# KINETIC SURFACE FRICTION RENDERING FOR INTERACTIVE SONIFICATION: AN INITIAL EXPLORATION

*Staas de Jong*

LIACS, Leiden University
Niels Bohrweg 1, Leiden
staas@liacs.nl

## ABSTRACT

Inspired by the role sound and friction play in interactions with everyday objects, this work aims to identify some of the ways in which kinetic surface friction rendering can complement interactive sonification controlled by movable objects. In order to do this, a tactile system is presented which implements a movable physical object with programmable friction. Important aspects of this system include the capacity to display high-resolution kinetic friction patterns, the ability to algorithmically define interactions directly in terms of physical units, and the complete integration of audio and tactile synthesis.

A prototype interaction spatially mapping arbitrary 1D signal data on a surface and directly converting these to sound and friction during movements across the surface is described. The results of a pilot evaluation of this interaction indicate how kinetic surface friction rendering can be a means for giving dynamically created virtual objects for sonification a tangible presence. Some specific possible roles for movement input and friction output are identified, as well as issues to be considered when applying and further developing this type of haptic feedback in the context of interactive sonification.

## 1. INTRODUCTION

In our everyday experience, friction and sound are often related phenomena. For example, when doing the dishes after a pleasant evening with friends, at some point we may find ourselves turning a piece of cloth over the surface of a wet plate. In what could be deemed a natural form of sonification, the presence or absence of squeaky sounds, in combination with the perceived smoothness of our movements, guides us in completing the task of making the surface dry. More generally, in any number of situations where we are moving some object over some surface, the resulting sound and friction will tell us something about both. When sound and friction can each be generated artificially as a means for display, this motivates the question how they can be artificially made to meaningfully complement eachother as well.

In the field of interactive sonification, this question may be relevant to systems where movable objects are used to control the way sound interactively displays data. For example, systems using computer mouse movement, or systems using tangible objects placed on a surface, such as presented in [2]. The goal of the initial exploration presented here is to identify ways in which kinetic surface friction rendering (the rendering of surface friction during movement) could complement such approaches to interactive sonification. To investigate this, we will use a prototype interaction which spatially maps an arbitrary one-dimensional amplitude series across a surface, and then directly converts it to sound and friction during movements over the surface. Here, physical exploration corresponds in a straightforward way to exploration of an underlying data space. Views at different levels of detail can be acquired, then, both by varying the spatial resolution of the mapping and by varying the movement speed at which the mapping is explored. In the prototype interaction, haptic friction output has been made an extension of sonic output in the sense of being concurrently and directly derived from it.

In the next section, the technology developed to provide kinetic surface friction rendering will be discussed first. After that, the prototype interaction's mechanism of converting spatially distributed signal data and movement input to real-time sound and friction output is described in detail. This is followed by the report on a pilot evaluation that was performed using the mechanism. The results of the evaluation are then discussed, and based on this the paper ends with conclusions and future work.

## 2. A SYSTEM FOR KINETIC SURFACE FRICTION RENDERING

The tactile interface of the proposed system consists of a freely movable object on a flat surface, pushed around by the user using the fingertips of one hand (see Figure 1). Inside the object is an electromagnet, mounted in a configuration putting its north- and south poles close together near the surface. The surface contains a ferromagnetic layer, so that a regulated vertical attraction between magnet and surface results in a variable horizontal friction during movement. The object's displacement is tracked using optical computer mouse hardware, while simultaneous auditory feedback is received via headphones.

In the past, a number of haptic computer mice have been proposed which used the same operating principle to generate friction. In [1], a friction mouse was presented, having a limited force range and 1-bit amplitude resolution. In [6], a friction mouse with an increased output force range was proposed, with friction controlled indirectly via an intermediate voltage range. In both cases, the intention was to provide the computer mouse with friction feedback in order to better support everyday Graphical User Interface (GUI) interactions. Within this context, the proposed devices were evaluated for target selection and pointing tasks.

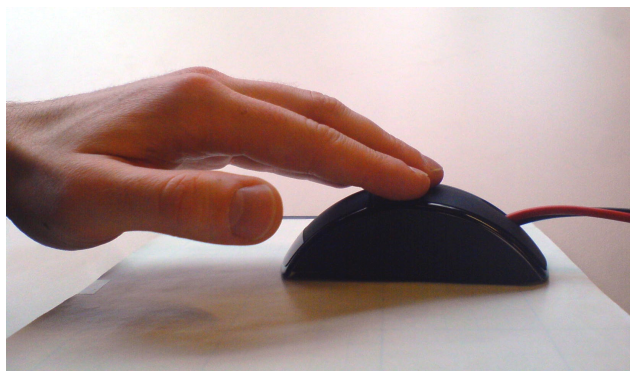

Figure 1. *The tactile interface.*

Instead of this, the goal of the current system has been to create a separate physical object with programmable friction, capable of displaying high-resolution patterns in kinetic friction. For one thing, this means that GUI screen coordinates are not used as a means for position input, since these are usually heavily influenced by intermediate velocity transformations and the artificial limitation to on-screen pixel display positions. Avoiding this, the object's physical position is tracked by directly accessing displacement input and, after conditioning, converting the updates to a signal in millimeters. Currently, this is done based on an update rate of 125 Hz, with the input sensitive to displacements down to 0.02 mm. The input speed signal derived simultaneously ranges from -334 to +334 mm/s, based on 255 input levels 2.6 mm/s apart.

The underlying goal of the system is also reflected in its output side. Using a method of force transfer which does not involve moving mechanical linkages holds the promise of fluid and precise force output. In order to explore this advantage, a custom electromagnet and current control circuit were developed. This has resulted in a kinetic friction range between 0.14 and 1.4 N, which seemed suitable for an object pushed around by the fingers. Friction amplitude resolution is limited only by the underlying analog electronics, and a smooth top layer has been added to the movement surface in order to have as little as possible of more subtle friction patterns drown in the tactile noise of normal operation. The audio-rate force output signal which the system provides is defined directly in terms of Newtons, with arbitrary temporal features across the previously mentioned output range programmable for durations down to 1 ms.

In order to support the full integration of audio and tactile synthesis, the system's tactile I/O has been implemented as a class in the SuperCollider language. In combination with the above, this supports algorithmically expressing interactions in terms of physical units such as Newtons, seconds and millimeters. In this way, interactions and insights gained about them can be more easily abstracted away from the actual technology and hardware being used.

### 3. CONVERTING SPATIALLY MAPPED SIGNAL DATA TO SOUND AND FRICTION

In one implemented mechanism for spatially mapping signal data, the signal fragment to be explored is first stored in a memory buffer, as a series of samples with arbitrary amplitude values between -1 and +1. From the displacement updates received from the device a continuous movement trajectory is reconstructed at a higher temporal and spatial resolution, based on linearly interpolating over the current update interval. The resulting signal is used, in combination with a variable parameter defining the number of samples per mm, to index into the signal buffer at audio rate. Using cubic interpolation, this directly generates vectors for audio playback.

To make the friction output a direct extension of sonic output, a sample-by-sample conversion of the latter to the former is used. First, for current sample index $t$ the average absolute amplitude $a_\mu[t]$ over the most recent millisecond of audio output is computed:

$$a_\mu[t] = \frac{1}{n} \sum_{i=0}^{i=n-1} |a[t-i]| \quad (1)$$

with $n$ corresponding to a period of 1 ms (here, $n = 192$). Then, in order to have friction force roughly match the perceived loudness of sonic output (rather than its numerical amplitude), the next friction value $f[t]$ is computed via:

$$f[t] = f_{mapping}(20 \log_{10}(\frac{a_\mu[t]}{a_{ref}})) \quad (2)$$

with $a_{ref} = 1$ so that the maximum absolute amplitude corresponds to 0 dB input to the $f_{mapping}$ function, which clips and linearly maps the range of [-30, 0] dB to [0.14, 0.5] N.

Because the system's audio output latency was measured as 1.4 ms, and its tactile output latency as 2.4 ms, audio output is delayed by 1.0 ms for better synchronization of sonic and tactile feedback.

In Figure 2, a recording is shown of movement speed, audio output and kinetic friction output during 5 seconds of interaction with a signal fragment using the above mechanism. In the left section of the figure, a signal feature is first explored in one direction; then, in the middle section, this is done in the opposite direction; and finally, with movement having returned more or less to the initial position, in the rightmost section the feature is again explored but now using a slower movement.

### 4. PILOT EVALUATION

A pilot experiment was conducted using 5 volunteer test subjects (2 males, 3 females) between 27 and 31 years old. None of the participants had any relevant hearing or manual impairments. The equipment used included the surface friction device described above, to which the test subjects were new. For audio output, a Motu UltraLite mk3 interface was used, operating at a 192 kHz sample rate. Test subjects listened to audio output via Beyerdynamic DT 770 M circumaural closed headphones, which isolated them from ambient noise, while providing a frequency range from 5 Hz to 30 kHz.

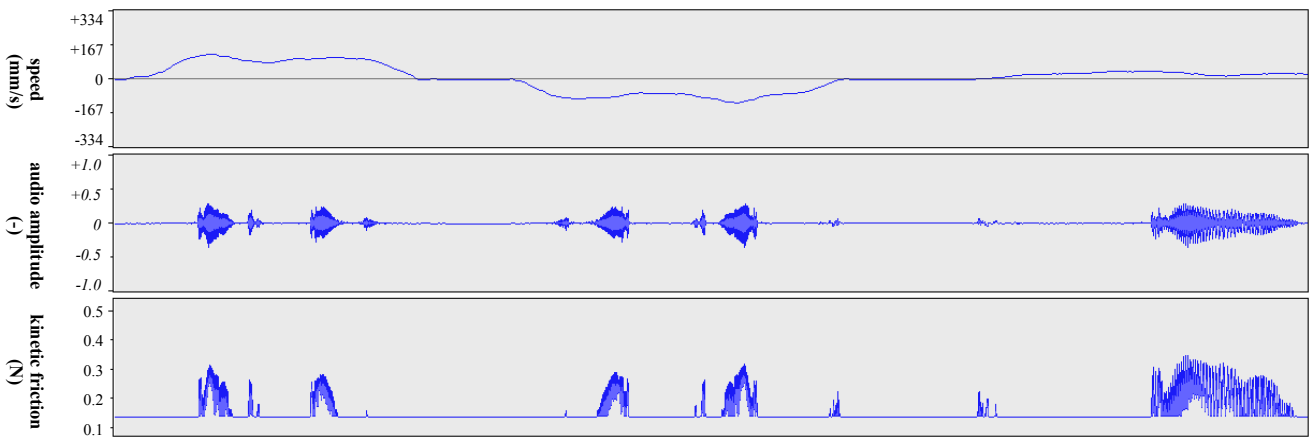

Figure 2. *A 5-second recording of movement speed, audio output and kinetic friction output during the spatial exploration of a signal feature.*

### 4.1. Procedure

Test subjects first received a spoken general introduction to the friction device, during which they were demonstrated how to move it around while touching it only with the tips of the middle three fingers of their hand of choice. They were then told they would be exploring a number of simple patterns, spread out sideways over the surface. These patterns would be represented in both sound and touch, and for each pattern they were invited to explore the surface using speeds from quite slow to fast.

The test subjects were then presented with a series of 3 different spatial mappings of the same signal fragment. In each case, this was done twice: first while friction output was turned off in the software, and then while it was turned on. The signal fragment would begin on the surface at the initial position of the device, and extend to the right of this. (Outside of this range, there was no audio or friction feedback.) The signal fragment itself consisted of a sinusoid of maximum amplitude, repeated 240 times with 100 samples for each cycle. Mappings 1, 2 and 3 are characterized in Table 1.

|  | samples per mm | sine cycle width | total signal fragment width | audio output frequency range during exploration |
|---|---|---|---|---|
| mapping 1 | 4000 | 0.025 mm | 6 mm | 104 - 13360 Hz |
| mapping 2 | 500 | 0.2 mm | 48 mm | 13 - 1670 Hz |
| mapping 3 | 8 | 12.5 mm | 3 m | 0.2 - 26.7 Hz |

Table 1. *Characterization of the different mappings used. (For mapping 3, only the first part of the signal fragment would fit on the physical surface used.)*

For the initial audio-only version of each mapping, test subjects were asked to characterize the audio sensations generated by describing them verbally. For the subsequent audio-haptic version they were asked to do the same, but now for the touch sensations. They were then asked to describe how they felt that sound and touch did or did not relate to eachother. When not exploring the full space or speed range, test subjects would receive verbal feedback from the experimenter (listening in on the audio output using a second pair of headphones) so that they could correct this. The ending criterium for each stage of the experiment would be the test subject indicating that he/she was done exploring, and satisfied with the completeness of the answers given.

### 4.2. Results

For mappings 1 and 2, all test subjects reported that the sound was "located" at a specific area of the surface, and that within this area their movement speed determined the sound's pitch. When friction feedback was added, all test subjects felt that the resulting tactile force sensations were located at a specific area; and also that sound and touch were located at or around the same spot. The resulting sensations were usually described in terms of "getting stuck" or a "syrupy" movement. For mapping 2, 4 subjects noted that slower movements yielded a regular pattern of change in the tactile sensation not present at higher speeds.

For mapping 3, none of the test subjects noticed the sine wave audio output, apart from an unintended artefact occurring when movement crossing into the signal fragment area activated signal playback. When friction feedback was activated, all test subjects felt that this sound artefact delineated where touch feedback began. Otherwise, the tactile feedback was felt to be unrelated to sound. The resulting sensation was characterized as that of sensing regular bumps on a surface by 4 of the test subjects, and described as that of "a gear with regular cogs, moving over the surface" by the remaining test subject.

However, one test subject likening the tactile sensation to bumps on a surface later corrected herself, stating it was somehow more like turning a rotary audio equipment knob with stepwise counterforces, since "overcoming one force peak, you end up at the next one, while with bumps you would normally get stuck in the valley inbetween".

## 5. DISCUSSION

The series of mappings presented during the experiment effectively let test subjects spatially zoom in into the microstructure of one and the same signal fragment. (Although they had not been told of this.) At the macro level corresponding to mapping 1 and, during fast movements, mapping 2, the tactile representation of spatially traversing a series of signal cycles being played back as audio was based on a heightened friction level computed from averaged signal intensity. Here, test subjects clearly felt that sound and touch were related, both resulting from movement in or over the same specific surface area. At the micro level corresponding to mapping 3, test subjects did not notice the audio rendering of signal cycles when moving within the signal fragment area, presumably because here output was largely infrasonic. They did clearly perceive the simultaneous tactile rendering, which at this level of mapping presented a somewhat arbitrary representation of microstructure, with two force cycles for each signal cycle (see Figure 3).

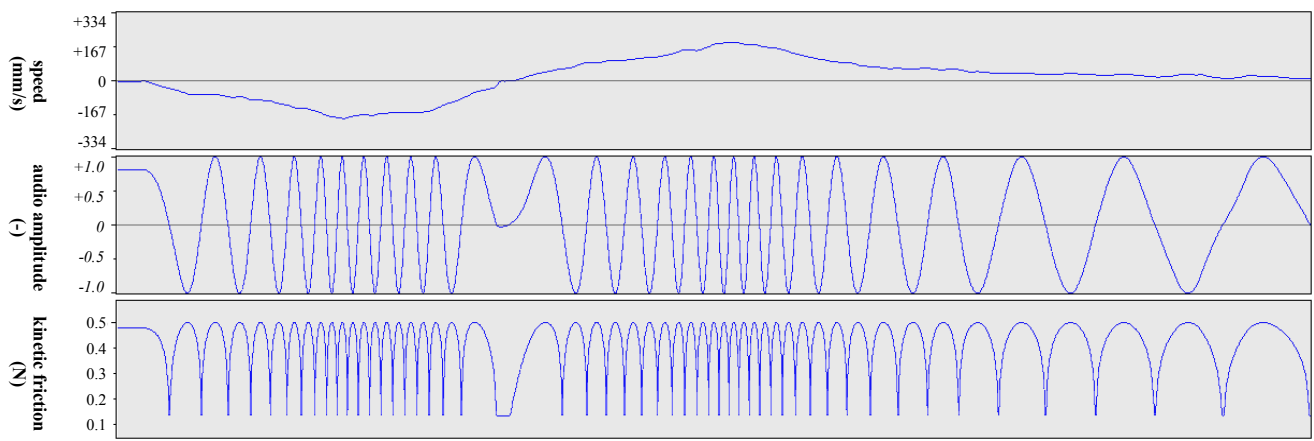

Figure 3. *A 4-second recording of movement speed, audio output and kinetic friction output during exploration of the pilot experiment signal fragment using spatial mapping 3.*

These force features were mostly perceived as regular bumps on the surface. This is reminiscent of how horizontal-only forces were used to create the sensation of vertical bumps in [5], verified for higher spatial frequencies for the device presented in [4]. However, if some other type of mapping had been used, the tactile representation of the microstructure could well have been perceived in terms of other sensations than such surface *unevenness*. For example, in [3] alternating regions of high and low resistance to movement are used to create a varying sense of surface *roughness*.

Although the output force range had been limited for the experiment, test subjects often remarked on the effects it had on their input. This included for example remarks on how "getting stuck" in the enlarged signal area of mapping 2 would bind them to slower movements. This illustrates that apart from being a means for additional display, kinetic friction can also be a means for directly influencing the movement soliciting display in general.

## 6. CONCLUSIONS AND FUTURE WORK

Our original goal was to identify ways in which kinetic surface friction rendering could complement interactive sonification controlled by movable objects on a surface. To this end, we presented a device tracking the 2D movement of an object on a surface, while providing a regulated attracting force between this object and the surface underneath. With the object's horizontal movement controlled by the fingertips, this was used as the tactile interface to a system implementing a separate physical object with programmable friction. Important aspects of this system include the capacity to display high-resolution kinetic friction patterns, the ability to algorithmically define interactions directly in terms of physical units, and the complete integration of audio and tactile synthesis.

The system was then used to implement a mechanism for spatially mapping arbitrary 1D signal data on a surface, with movements across the surface generating a sonic and frictional readout. Important characteristics of this mechanism included friction being a direct extension of sonic output; and the ability to provide a differentiated display of signal macro- and microstructure. This is done both by varying the resolution of the spatial mapping, and by varying the movement speed at which it is explored during interaction. The prototype interaction was evaluated in a pilot experiment where test subjects explored an example signal fragment, with and without haptic feedback, at different movement speeds, and at different spatial mapping resolutions.

The results indicate that kinetic surface friction rendering can be used as a means for giving dynamically created virtual objects for sonification a tangible presence. Here, the speed of physical movement can be a straightforward and intuitive way of controlling the level of detail at which data is explored. Sonic and tactile features mapped to the same area were indeed perceived as belonging to the same location, suggesting the use of friction for spatial orientation. Also, the perception of infrasonic signals as a bumpy surface suggests the potential use of kinetic friction as a type of spectral extension to sonification, displaying frequencies too low to be made heard. However, this also illustrates how a seemingly general conversion from sound to friction has resulted in a quite specific type of tactile sensation – which might have been qualitatively different had some other conversion been used.

And on another note, already within the limited force range used in the experiment, changes in friction level were clearly able to significantly slow down or speed up the controlling movements. In this way, adding haptic feedback not only meant adding an additional channel of tactile display; it also created a means of directly influencing interaction by altering the movement navigating the display.

Clearly, the mechanism that was experimented with here is only one of many possible ways to meaningfully combine sound and friction for interaction. In future work, the spatiotemporal friction patterns to be used should probably first be explicitly considered and explored for the types of tactile sensations they induce. The insights gained by doing this are expected to enable more refined and intuitive re-couplings of the two modalities. Another question that should be explicitly considered when designing future interactions is whether kinetic friction is intended purely as a means for tactile display during movement, or also as a means to directly influence movement input and thereby the resulting sonic exploration.

## 7. ACKNOWLEDGMENTS

The author would like to thank Rene Overgauw at the Electronics Department of the Faculty of Science of Leiden University for his indispensable advice and practical support. Also thanks to all test subjects for kindly volunteering, and to the ISon2010 Program Committee for making the consideration of this contribution possible.